\documentclass[conference]{IEEEtran}
\usepackage{url,array,lipsum,hhline,graphicx,subfigure,amsfonts,cite,epstopdf,makeidx,outlines,textcomp,multirow,soul,hyperref,enumitem,amssymb,amsmath,color,graphicx, amsbsy,}
\usepackage[table,xcdraw]{xcolor}
\usepackage[linesnumbered,ruled]{algorithm2e}
\usepackage[noprefix]{nomencl}
\usepackage[none]{hyphenat}
\allowdisplaybreaks

\begin{document}

\title{Strategic Competition of Electric Vehicle Charging Stations in a Regulated Retail Electricity Market}

\author{
\IEEEauthorblockN{1\textsuperscript{st} Reza Bayani}
\IEEEauthorblockA{\textit{ Electrical and Computer Engineering} \\
\textit{University of California, San Diego}\\
San Diego, USA 92161 \\
rbayani@ucsd.edu}
\and
\IEEEauthorblockN{2\textsuperscript{nd} Arash Farokhi Soofi}
\IEEEauthorblockA{\textit{  Electrical and Computer Engineering} \\
\textit{University of California, San Diego}\\
San Diego, USA 92161  \\
afarokhi@ucsd.edu}
\and
\IEEEauthorblockN{3\textsuperscript{rd} Saeed D. Manshadi}
\IEEEauthorblockA{\textit{ Electrical and Computer Engineering} \\
\textit{San Diego State University}\\
San Diego, USA 92182 \\
smanshadi@sdsu.edu}}

\maketitle

\bstctlcite{IEEEexample:BSTcontrol} 

\begin{abstract}
The increasing trend of transportation electrification presents investors the opportunity to provide charging services to Electric Vehicle (EV) owners via the energy purchased from the wholesale electricity market. 
This will benefit EV owners with the availability of competitive rates compared to the regulated utility time-of-use (TOU) rates. The fundamental questions addressed in this paper are \emph{1) will EV owners benefit from the additional choice of Electric Vehicle Charging Stations (EVCSs) compared to home charging? 2) is there any profitable market opportunity for charging stations while the retail electricity market is regulated?}  
To this end, the strategic bidding problem for EVCSs which purchase electricity from the Day-Ahead Electricity Market (DAM) and sell it to EV owners is presented. The strategic bidding problem is constrained by the market-clearing problem within the DAM as well as EVs' charging cost minimization problem. A bi-level optimization problem formulation and a solution method are presented to address this work's research questions. The effectiveness of the proposed structure in gaining profit for EVCSs is illustrated, and it is shown that EV owners also save on their charging cost with the presence of EVCSs as a choice.
\end{abstract}

\vspace{-.35cm}
\section*{Nomenclature}
\subsection*{Indices, Sets, and Superscripts}
\noindent \begin{tabular}{ll}
$b,\mathcal{B},B$ & Buses in the power system\\
$c,\mathcal{C},C$ & Charging stations\\
$ch/dc$ & Charge/discharge of EV fleets \\
$d,\mathcal{D},D$ & Electricity demands\\
$f,\mathcal{F},F$ & EV fleets\\
$g,\mathcal{G},G$ & Generation units\\
$k,\mathcal{K},K$ & Segments in generation cost function\\
$m,\mathcal{M},M$ & Segments in bidding price function\\
$l,\mathcal{L},L$ & Lines in the power system\\
$s,\mathcal{S},S$ & Solar photovoltaic units\\
$t,\mathcal{T}$ & Time steps\\
$to/fr$ & To/from\\
$\underline{*},\overline{*}$ & Lower/upper bound value indicators\\
\end{tabular}
\subsection*{Variables}
\noindent \begin{tabular}{ l p{6.55cm} }
$e$ & Energy level of EV fleet\\
$p$ & Real power\\
$\pi$ & Willingness to pay bid of charging stations\\
$\tau$ & Offering price of charging stations\\
$\theta$ & Voltage angle\\
$\mu, \lambda$ & Dual variables \\
\end{tabular}

\subsection*{Parameters}
\noindent \begin{tabular}{ l p{6.55cm} }
$x$ & Reactance of line\\
$\xi$ & Cost of generation\\
$\epsilon$ & EV fleet connectivity ratio\\
$\eta$ & Charge/discharge efficiency\\
$\kappa$ & Time-of-use rate\\
\end{tabular}

\section{Introduction}
\par Predictions on Electric Vehicle (EV) adoption rate are not in complete accord, and several organizations have reported EV adoption projections ranging from 10\% to 60\% by the year 2040 \cite{muratori2021rise}. However, it is very likely that with technological improvements, consumer awareness, and supporting policies, EV sales will soar. California has passed a bill that requires all new cars sold in the state to be zero-emission by the year 2035 \cite{ca2035zero}. 
Construction of more EVCSs is crucial to keep up with the increasing EV demand, manage and regulate the impact of EV charging on the power system, and promote public equity. Although transportation electrification is desirable from aspects such as greenhouse gas emission reduction \cite{bayani2021coordinated}, grid services \cite{bayani2021autonomous,bhusal2021cybersecurity}, and reducing energy costs \cite{babaei2021data}, EV charging without proper management  introduces several challenges to the power system \cite{soofi2020investigating,soofi2021analyzing}.
It has also been shown that neighborhoods with below-median household income are less likely to have access to residential or public EVCSs \cite{hsu2021public}. Motivated by these facts, this study aims to investigate the profitability of the operation of EVCSs in the power system.

\par Researchers have studied the expected growth in the integration of EVs and EVCSs into the power system from several aspects. 
EV participation in the market through learning algorithms is studied in \cite{najafi2019reinforcement}. A study on the participation of aggregated EVs in day-ahead energy and reserve markets is presented in \cite{khoshjahan2020optimal}. 
Optimal power flow in the presence of EVCSs has been discussed in \cite{fakouri2021active}.
A detailed study on the contributing factors in the economics of EVCSs can be found in \cite{zhang2018factors}. A control algorithm for operation cost reduction of EVCSs participating in the market is presented in \cite{yan2018optimized}. Participation of EVCSs in the flexibility market is studied in \cite{divshali2021optimum}. A privacy-preserving pricing strategy for profit maximization of EVCSs is presented in \cite{lee2021dynamic}. 

\par This work investigates the economic viability of developing EVCSs in the power system by considering social welfare and EV owner costs. We propose a market structure that enables EVCSs to purchase electricity in the day-ahead market (DAM) and offer it to customers. EVCSs submit Willing to Pay (WTP) bids for purchasing energy in the DAM. A bi-level model is proposed, which maximizes the profits of EVCSs in the upper level. At the same time, EV owners' electricity costs are minimized, and social welfare is maximized in the lower-level problem.

\section{Problem Formulation}
\begin{figure}[h!]
    \centering
    \vspace{-.35cm}    \includegraphics[width=1\linewidth]{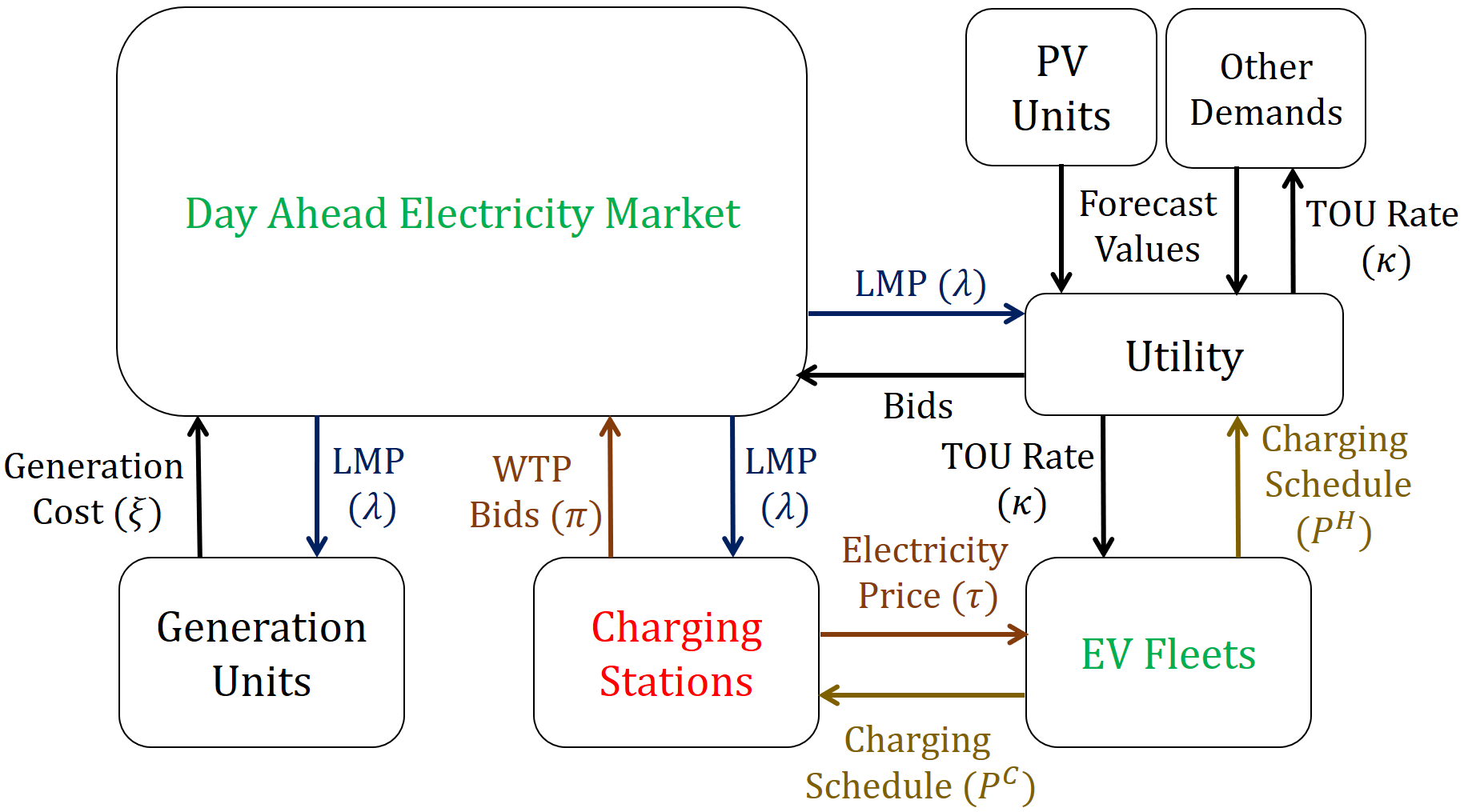}
    \vspace{-.65cm}
    \caption{Structure of the proposed bidding/offering problem}
    \label{fig:evcs_structure} \vspace{-.35cm}
\end{figure}

In this section, first, the schematic of the proposed structure is presented and discussed. The market structure developed for modeling the optimal bidding/offering strategy is illustrated in Fig. \ref{fig:evcs_structure}. The participants of the DAM consist of generation units, EVCSs, and the utility. Generation units submit their generation bids to the DAM, while EVCSs and the utility submit their WTP bids. EV fleets are provided with offers from both EVCSs and the regulated time-of-use (TOU) rates from the utility. From the EVCS point of view, the proposed structure is modeled by a bi-level formulation. EVCSs aim to maximize their collective profit in the upper-level problem. Two lower-level problems are considered which model the power system market and EV fleets' operation. EV fleets aim to choose the lowest cost charging strategy. The power system market operation is also modeled to maximize social welfare, which is defined as the overall WTP of EVCSs minus the total costs of generation units.

\par In the rest of this section, the proposed bi-level bidding/offering model is discussed in three parts.

\subsection{Profit maximization of the EVCSs }
It is supposed that several EVCSs are present at select buses of the network. In the upper level, EVCSs try to find the optimal WTP bids and offer prices so that their profit is maximized. \vspace{-.35cm}

\begin{subequations}\label{eq:evcs_upper}
\begin{alignat}{3} 
    &    \max  \textstyle \sum_{\mathcal{T}}\sum_{\mathcal{C}}\sum_{\mathcal{F}}{(p^C_{f,c,t}\cdot\tau_{f,c,t}-p^C_{f,c,t}\cdot\lambda^B_{\mathcal{B}_f,t})}\label{eq:evcs_obj_upper}\\
    & \nonumber \text{subject to:} \\
    & \underline{\tau}_{f,c,t} \leq \tau_{f,c,t}\leq \overline{\tau}_{f,c,t}, \hspace{1.cm} \forall \; \mathcal{F},\mathcal{C},\mathcal{T}\label{eq:evcs_offer_lims}
\end{alignat}\end{subequations}

\par The objective function of the upper-level problem is presented by Eq. \eqref{eq:evcs_obj_upper}, which maximizes the overall profits of EVCSs. This objective is calculated by subtracting EVCSs' purchase costs from their revenues. In this case, revenues are represented by the first term in \eqref{eq:evcs_obj_upper}, which equals the total power sold at charging stations multiplied by the offer price. The second term in \eqref{eq:evcs_obj_upper} models the purchase costs of EVCSs and is obtained through multiplying total power charged at EVCS locations by the Local Marginal Price (LMP) cleared by the DAM. The EVCS offer price is also constrained by \eqref{eq:evcs_offer_lims}.

\subsection{Day-Ahead Electricity Market Problem}
DAM constitutes one of the lower-level problems of the proposed model and is presented in Eq. \eqref{eq:evcs_powersys_primal}. In this problem, the aim is to maximize the social welfare, which is equivalent to subtracting generation costs from collective WTP of EVCSs, respectively modeled by the second and the first elements in the objective function Eq. \eqref{eq:evcs_powersys_obj}. It is assumed that generation units are bidding based on their marginal costs. That is why in the social welfare maximization problem, the generation units' bids are substituted with the cost of generation.

\vspace{-.15cm}
\begin{subequations}\label{eq:evcs_powersys_primal}
\begin{alignat}{3}
& \begin{aligned}
    \max  \textstyle \sum_{\mathcal{F}}\sum_{\mathcal{C}}\sum_{\mathcal{M}}\sum_{\mathcal{T}}{p^M_{f,c,m,t}\cdot \pi_{f,c,m,t}} \\
    -  \textstyle \sum_{\mathcal{G}}\sum_{\mathcal{K}}\sum_{\mathcal{T}}{\xi_{g,k}\cdot p^K_{g,k,t}} \label{eq:evcs_powersys_obj}\end{aligned}\\
& \nonumber \text{subject to:}\\
& \begin{aligned} \underline{\pi}_{f,c,m,t} \leq \pi_{f,c,m,t}\leq \overline{\pi}_{f,c,m,t}, \hspace{3.5cm} \\  \forall \; \mathcal{F},\mathcal{C},\mathcal{M},\mathcal{T}  : \underline{\mu}^P_{f,c,m,t},\overline{\mu}^P_{f,c,m,t} \label{eq:evcs_powersys_wtp_limit}\end{aligned}\\
& \underline{p}^G_g \leq p^G_{g,t} \leq \overline{p}^G_g,&\hspace{-5.8cm} \forall\; \mathcal{G},  \mathcal{T} : \underline{\mu}^G_{g,t},\overline{\mu}^G_{g,t} \label{eq:evcs_powersys_p_g_limit}\\
& \underline{p}^L_{l} \leq p^L_{l,t}\leq \overline{p}^L_{l} ,&\hspace{-5.8cm} \forall \; \mathcal{L}, \mathcal{T}  : \underline{\mu}^L_{l,t},\overline{\mu}^L_{l,t} \label{eq:evcs_powersys_p_l_limit}\\
& \underline{\theta}_{b} \leq \theta_{b,t}\leq \overline{\theta}_{b}, &\hspace{-5.8cm} \forall \; \mathcal{B}, \mathcal{T}  : \underline{\mu}^B_{b,t},\overline{\mu}^B_{b,t} \label{eq:evcs_powersys_theta_limit}\\
& 0 \leq p^{S}_{s,t} \leq \overline{p}^{S}_{s,t},&\hspace{-5.8cm} \forall \; \mathcal{S}, \mathcal{T} : \underline{\mu}^S_{s,t},\overline{\mu}^S_{s,t} \label{eq:evcs_powersys_p_s_limit}\\
& \underline{p}^K_{g,k} \leq p^K_{g,k,t} \leq \overline{p}^K_{g,k},&\hspace{-5.8cm} \forall \; \mathcal{G},  \mathcal{K},\mathcal{T}  : \underline{\mu}^K_{g,k,t},\overline{\mu}^K_{g,k,t} \label{eq:evcs_powersys_p_g_k_limit}\\
& p^G_{g,t} =  \textstyle \sum_{\mathcal{K}}{p}^K_{g,k,t},&\hspace{-5.8cm} \forall \; \mathcal{G}, \mathcal{T}  : {\lambda}^G_{g,t} \label{eq:evcs_powersys_p_g_k_eq}\\
& p^L_{l,t}  = \left(\theta_{\mathcal{B}^{fr}_{l},t}-\theta_{\mathcal{B}^{to}_{l},t}\right)/x_{l}, &\hspace{-5.8cm} \forall  \mathcal{L}, \mathcal{T}  : {\lambda}^L_{l,t} \label{eq:evcs_powersys_dc_pf}\\
&\begin{aligned}
 \textstyle \sum_{\mathcal{G}_b}^{}{p^G_{g,t}}+\sum_{\mathcal{S}_b}^{}{p^{S}_{s,t}}-\sum_{\mathcal{L}_b^{fr}}{p^L_{l,t}}+\sum_{\mathcal{L}_b^{to}}{p^L_{l,t}}\hspace{1.35cm}\\  \textstyle =\sum_{\mathcal{D}_b}{p^D_{d,t}}+\sum_{\mathcal{F}_b}{p^F_{f,t}}, \hspace{1.05cm}\forall \mathcal{B}, \mathcal{T} : {\lambda}^B_{b,t} \end{aligned}\label{eq:evcs_powersys_nodal_balance}
\end{alignat}
\end{subequations}

\par The rest of the DAM problem constraints include limits for WTP bids, generation of units, thermal capacity of electricity lines, voltage angle bounds, and solar units' dispatch, which are presented by Eqs. \eqref{eq:evcs_powersys_wtp_limit} through \eqref{eq:evcs_powersys_p_s_limit}. Constraints \eqref{eq:evcs_powersys_p_g_k_limit} and \eqref{eq:evcs_powersys_p_g_k_eq} represent the piecewise linear function with which the generation cost model is presented. DC power flow utilized for the market operation and is presented in Eq. \eqref{eq:evcs_powersys_dc_pf}. Finally, Eq. \eqref{eq:evcs_powersys_nodal_balance} presents the nodal electricity balance at all buses and times. In front of all the presented constraints, their respective dual variables are shown. LMP at each node and time is equal to the dual value of constraint \eqref{eq:evcs_powersys_nodal_balance}.

\subsection{Optimal Operation Problem of EV Fleets }
The governing constraints of the optimal EV fleet operation problem are presented in Eq. \eqref{eq:evcs_fleet_primal}. EV fleets aim to minimize their aggregated electricity charging costs. EVs can either be charged at charging stations or their home (indexed by \textit{H}). In the former case, EVs are charged by offer price of EVCSs (the first term in \eqref{eq:evcs_fleet_obj}) and in the latter case, they are charged by TOU rate (the second term in \eqref{eq:evcs_fleet_obj}).
\vspace{-.15cm}
\begin{subequations}\label{eq:evcs_fleet_primal}
\begin{alignat}{3}
& \begin{aligned}
\textstyle \min \sum_{\mathcal{F}}\sum_{\mathcal{C}}\sum_{\mathcal{T}}{p^C_{f,c,t}\cdot \tau_{f,c,t}}+\sum_{\mathcal{F}}\sum_{\mathcal{T}}{p^H_{f,t}\cdot\kappa_{f,t}} \label{eq:evcs_fleet_obj}\end{aligned}\hspace{.45cm}\\
& \nonumber \text{subject to:}\\
& 0 \leq p^F_{f,t} \leq \overline{p}^F_{f},&\hspace{-6.95cm} \forall\; \mathcal{F}, \mathcal{T}  : \underline{\mu}^{F}_{f,t},\overline{\mu}^{F}_{f,t} \label{eq:evcs_p_f_limit}\\
& 0 \leq p^C_{f,c,t} \leq \epsilon^{C}_{f,c,t}\cdot\overline{p}^C_{f,c},&\hspace{-6.95cm} \forall\; \mathcal{F},\mathcal{C}, \mathcal{T}  : \underline{\mu}^{C}_{f,c,t},\overline{\mu}^{C}_{f,c,t} \label{eq:evcs_p_c_limit}\\
& 0 \leq p^H_{f,t} \leq \epsilon^{H}_{f,t}\cdot\overline{p}^H_{f},&\hspace{-6.95cm} \forall\; \mathcal{F}, \mathcal{T}  : \underline{\mu}^{H}_{f,t},\overline{\mu}^{H}_{f,t} \label{eq:evcs_p_h_limit}\\
& \textstyle p^F_{f,t} = p^H_{f,t} + \sum_{\mathcal{C}}^{}{p^C_{f,c,t}} ,&\hspace{-6.95cm} \forall\; \mathcal{F}, \mathcal{T}  : \lambda^{F}_{f,t} \label{eq:evcs_p_f_eq}\\
& 0 \leq p^M_{f,c,m,t} \leq \overline{p}^M_{f,c,m},&\hspace{-6.95cm} \forall\; \mathcal{F},\mathcal{C},\mathcal{M},\mathcal{T}  : \underline{\mu}^{M}_{f,c,m,t},\overline{\mu}^{M}_{f,c,m,t} \label{eq:evcs_p_c_m_limit}\\
& \textstyle p^C_{f,c,t} = \sum_{\mathcal{M}}^{}{p^M_{f,c,m,t}},&\hspace{-6.95cm} \forall\; \mathcal{F},\mathcal{C}, \mathcal{T}  : \lambda^{C}_{f,c,t} \label{eq:evcs_p_c_eq}\\
& \underline{e}^F_{f} \leq e^F_{f,t} \leq \overline{e}^F_{f},&\hspace{-6.95cm} \forall\; \mathcal{F}, \mathcal{T}: \underline{\mu}^E_{f,t},\overline{\mu}^E_{f,t}\label{eq:evcs_e_f_limit}\\
& e^F_{f,t+1}= e^F_{e,t}- \frac{p^{dc}_{f,t}}{\eta^{dc}_{e}} +p^{F}_{f,t}\cdot\eta^{ch}_{e} ,&\hspace{-6.95cm} \forall\; \mathcal{F}, \mathcal{T} : {\lambda}^E_{f,t}\label{eq:evcs_e_f_eq}
\end{alignat}
\end{subequations}
According to \eqref{eq:evcs_p_f_limit}, the overall charged power of each fleet is limited the maximum charging rate of the batteries. By Eq. \eqref{eq:evcs_p_f_eq}, EVs are connected at EVCS locations or are plugged at home. The charging power bounds of EV fleets at EVCSs and homes are displayed by Eqs. \eqref{eq:evcs_p_c_limit} and \eqref{eq:evcs_p_h_limit}, respectively.
The piecewise linear function for modeling the WTP bids of EVCSs are displayed by \eqref{eq:evcs_p_c_m_limit} and \eqref{eq:evcs_p_c_eq}. Finally, the energy capacity of each fleet at each time is limited by \eqref{eq:evcs_e_f_limit} and the energy equation is represented by \eqref{eq:evcs_e_f_eq}.

\section{Solution Methodology}
The presented formulations in the previous section form a bi-level program of the following general form:
\begin{equation} \label{eq:evcs_bi_level}
    \hspace{-.5cm}\max f(x),\; s.t: x\in \arg\max g(x), x\in  \arg\min h(x), x\in\mathcal{X}
\end{equation}
where \textit{f(x)}, \textit{g(x)}, and \textit{h(x)} respectively represent the objectives in Eqs. \eqref{eq:evcs_upper}, \eqref{eq:evcs_powersys_primal}, and \eqref{eq:evcs_fleet_primal}, and the feasible region is represented by $\mathcal{X}$. In order to solve \eqref{eq:evcs_bi_level}, we first obtain the dual forms of \eqref{eq:evcs_powersys_primal} and \eqref{eq:evcs_fleet_primal}. The dual problem of \eqref{eq:evcs_powersys_primal} is presented in \eqref{eq:evcs_powersys_dual} as follows:
\begin{subequations}\label{eq:evcs_powersys_dual}
\begin{alignat}{3}
& \begin{aligned}
\min \textstyle \sum_{\mathcal{G}}\sum_{\mathcal{K}}\sum_{\mathcal{T}}( \underline{\mu}^K_{g,k,t}\cdot\underline{p}^K_{g,k}
-\overline{\mu}^K_{g,k,t}\cdot\overline{p}^K_{g,k}) \hspace{2.2cm}\\ +
\textstyle \sum_{\mathcal{B}}\sum_{\mathcal{T}}( \lambda^B_{b,t}(\sum_{\mathcal{D}_b}p^D_{d,t}+\sum_{\mathcal{F}_b}p^F_{f,t})+\underline{\mu}^B_{b,t}\cdot\underline{\theta}_{b}
-\overline{\mu}^B_{b,t}\cdot\overline{\theta}_{b})\\
+ \textstyle \sum_{\mathcal{L}}\sum_{\mathcal{T}}( \underline{\mu}^L_{l,t}\cdot\underline{p}^L_{l}
-\overline{\mu}^L_{l,t}\cdot\overline{p}^L_{l}) -  \textstyle \sum_{\mathcal{S}}\sum_{\mathcal{T}}
\overline{\mu}^S_{s,t}\cdot\overline{p}^S_{s,t}  \\ +
\textstyle \sum_{\mathcal{F}}\sum_{\mathcal{C}}\sum_{\mathcal{M}}\sum_{\mathcal{T}}{(\underline{\mu}^P_{f,c,m,t}\cdot\underline{\pi}_{f,c,m,t} -\overline{\mu}^P_{f,c,m,t}\cdot\overline{\pi}_{f,c,m,t})}\\ +
 \textstyle \sum_{\mathcal{G}}\sum_{\mathcal{T}}{( \underline{\mu}^G_{g,t}\cdot\underline{p}^G_{g} -\overline{\mu}^G_{g,t}\cdot\overline{p}^G_{g})}   \end{aligned} \label{eq:evcs_powersys_dual_obj}
\end{alignat}
\vspace{-.85cm}
\end{subequations}
\setcounter{equation}{4}
\begin{subequations}
\setcounter{equation}{1}
\begin{alignat}{3}
& \nonumber \text{subject to:}\\
& \underline{\mu}^P_{f,c,m,t}-\overline{\mu}^P_{f,c,m,t} = p^M_{f,c,m,t}, &\hspace{-1.59cm} \forall  \mathcal{F},\mathcal{C},\mathcal{M},\mathcal{T}  : {\pi}_{f,c,m,t} \label{eq:evcs_powersys_dual_begin}\\
&  \textstyle \underline{\mu}^G_{g,t}-\overline{\mu}^G_{g,t} + {\lambda}^G_{g,t} +  \sum_{\mathcal{G}_b}{\lambda}^B_{b,t} = 0,& \hspace{-1.59cm} \forall\; \mathcal{G},  \mathcal{T} : p^G_{g,t} \\
& \underline{\mu}^K_{g,k,t}-\overline{\mu}^K_{g,k,t} - {\lambda}^G_{g,t} = -\xi_{g,k},& \hspace{-1.59cm} \forall\; \mathcal{G}, \mathcal{K}, \mathcal{T} : p^K_{g,k,t} \\
&  \textstyle  \underline{\mu}^S_{s,t}-\overline{\mu}^S_{s,t} + \sum_{\mathcal{S}_b}{\lambda}^B_{b,t} = 0,& \hspace{-1.59cm} \forall\; \mathcal{S},  \mathcal{T} : p^S_{s,t} \\
&  \textstyle \underline{\mu}^B_{b,t}-\overline{\mu}^B_{b,t} + \sum_{\mathcal{L}^{to}_b}{\dfrac{\lambda^L_{l,t}}{x_{l}}} = \sum_{\mathcal{L}^{fr}_b}{\dfrac{\lambda^L_{l,t}}{x_{l}}} ,& \hspace{-1.59cm} \forall \; \mathcal{B}, \mathcal{T}  : \theta_{b,t} \\
&  \textstyle \underline{\mu}^L_{l,t}-\overline{\mu}^L_{l,t} + {\lambda}^L_{l,t} + \sum_{\mathcal{L}^{to}_b}{{\lambda}^B_{b,t}}
= \sum_{\mathcal{L}^{fr}_b}{{\lambda}^B_{b,t}} ,&\hspace{-1.59cm} \forall \; \mathcal{L}, \mathcal{T}  : p^L_{l,t}\\
& \mu \leq 0, \lambda \label{eq:evcs_powersys_dual_end}
\end{alignat}
\end{subequations}

\par Eq. \eqref{eq:evcs_fleet_dual} captures the dual form of \eqref{eq:evcs_fleet_primal}:
\vspace{-.25cm}
\begin{subequations}\label{eq:evcs_fleet_dual}
\begin{alignat}{3}
    & \begin{aligned}
    \textstyle \max \sum_{\mathcal{T}}\sum_{\mathcal{F}}(- \overline{\mu}^{F}_{f,t} \cdot \overline{p}^F_{f}
 - \frac{p^{dc}_{f,t}}{\eta^{dc}_{e}}\cdot{\lambda}^E_{f,t} \hspace{1.5cm}\\ + \underline{\mu}^E_{f,t}\cdot\underline{e}^F_{f} - \overline{\mu}^E_{f,t}\cdot\overline{e}^F_{f} - \overline{\mu}^{H}_{f,t}\cdot\epsilon^{H}_{f,t}\cdot\overline{p}^H_{f}\\ \textstyle - \sum_{\mathcal{C}} \epsilon^{C}_{f,c,t}\cdot\overline{\mu}^{C}_{f,c,t}\cdot\overline{p}^C_{f,c} + \sum_{\mathcal{C}}\sum_{\mathcal{M}} \overline{\mu}^{M}_{f,c,m,t}\cdot\overline{p}^M_{f,c,m}) \end{aligned}\label{eq:evcs_fleet_dual_obj}\\
& \nonumber \text{subject to:}\\
&  \underline{\mu}^{F}_{f,t}-\overline{\mu}^{F}_{f,t}+\lambda^{F}_{f,t}-\lambda^{E}_{f,t}\cdot\eta^{ch}_{e}=0 ,&\hspace{-3.92cm} \forall\; \mathcal{F}, \mathcal{T}  :p^F_{f,t}\label{eq:evcs_fleet_dual_begin}\\
& \underline{\mu}^{C}_{f,c,t}-\overline{\mu}^{C}_{f,c,t} + \lambda^{C}_{f,c,t} - \lambda^{F}_{f,t} =0 ,&\hspace{-3.92cm} \forall\; \mathcal{F},\mathcal{C}, \mathcal{T}  : p^C_{f,c,t} \\
& \begin{aligned}\underline{\mu}^{M}_{f,c,m,t}-\overline{\mu}^{M}_{f,c,m,t}-\lambda^{C}_{f,c,t} = \pi_{f,c,m,t} ,\hspace{2.3cm} \\\forall\; \mathcal{F},\mathcal{C},\mathcal{M},\mathcal{T}  : p^M_{f,c,m,t}\end{aligned}\\
& \underline{\mu}^{H}_{f,t}-\overline{\mu}^{H}_{f,t}-\lambda^{F}_{f,t}=\kappa_{f,t},&\hspace{-3.92cm} \forall\; \mathcal{F}, \mathcal{T}  :p^H_{f,t} \\
& \underline{\mu}^E_{f,t}-\overline{\mu}^E_{f,t}+{\lambda}^E_{f,t-1}-{\lambda}^E_{f,t}=0,&\hspace{-3.92cm} \forall\; \mathcal{F}, \mathcal{T}:e^F_{f,t}\\
& \mu \geq 0, \lambda \label{eq:evcs_fleet_dual_end}
\end{alignat}
\end{subequations}
\par Now, the equivalent single-level form of \eqref{eq:evcs_bi_level} can be written with the help of the presented dual forms and the Slater's condition:
\begin{subequations} \label{eq:evcs_single_level}
\begin{alignat}{3}
    & \max \eqref{eq:evcs_obj_upper} \\
    & \nonumber \text{subject to:}\\
    & \nonumber \text{All of the constraints in \eqref{eq:evcs_upper},\;\eqref{eq:evcs_powersys_primal},\;\eqref{eq:evcs_fleet_primal},\;\eqref{eq:evcs_powersys_dual},\;\eqref{eq:evcs_fleet_dual}} \\
    & \eqref{eq:evcs_powersys_obj}=\eqref{eq:evcs_powersys_dual_obj}\\
    & \eqref{eq:evcs_fleet_obj}=\eqref{eq:evcs_fleet_dual_obj}
\end{alignat}
\end{subequations}
\par By forming the single-level problem in \eqref{eq:evcs_single_level}, the solution method is completed and the optimal bidding and offering problem can be solved. The final problem presented in \eqref{eq:evcs_single_level} is a non-linear optimization problem that can be solved by reliable off-the-shelf non-linear solvers. The results presented in the next section illustrate that the provided solution method is valid, i.e. the equality constraints in \eqref{eq:evcs_single_level} are satisfied and each sub-problem has a feasible region.

\section{Results and discussions} 
Several cases are illustrated by simulating the operation of the proposed structure in a modified IEEE 30-bus transmission network. It is assumed that 3 EVCSs are operating at each 10 buses of the network where EV fleets are present. EVs face a choice to be charged at either of these EVCSs or at home. In addition, 4 Photovoltaic (PV) units are also considered to be installed in the system. The required network data and parameters for replication of the results in the case studies are available at \cite{data_github}.

\subsection{Validation of the Proposed Structure}

\begin{figure}[h!]
    \centering
    \vspace{-.35cm}
    \includegraphics[width=\linewidth]{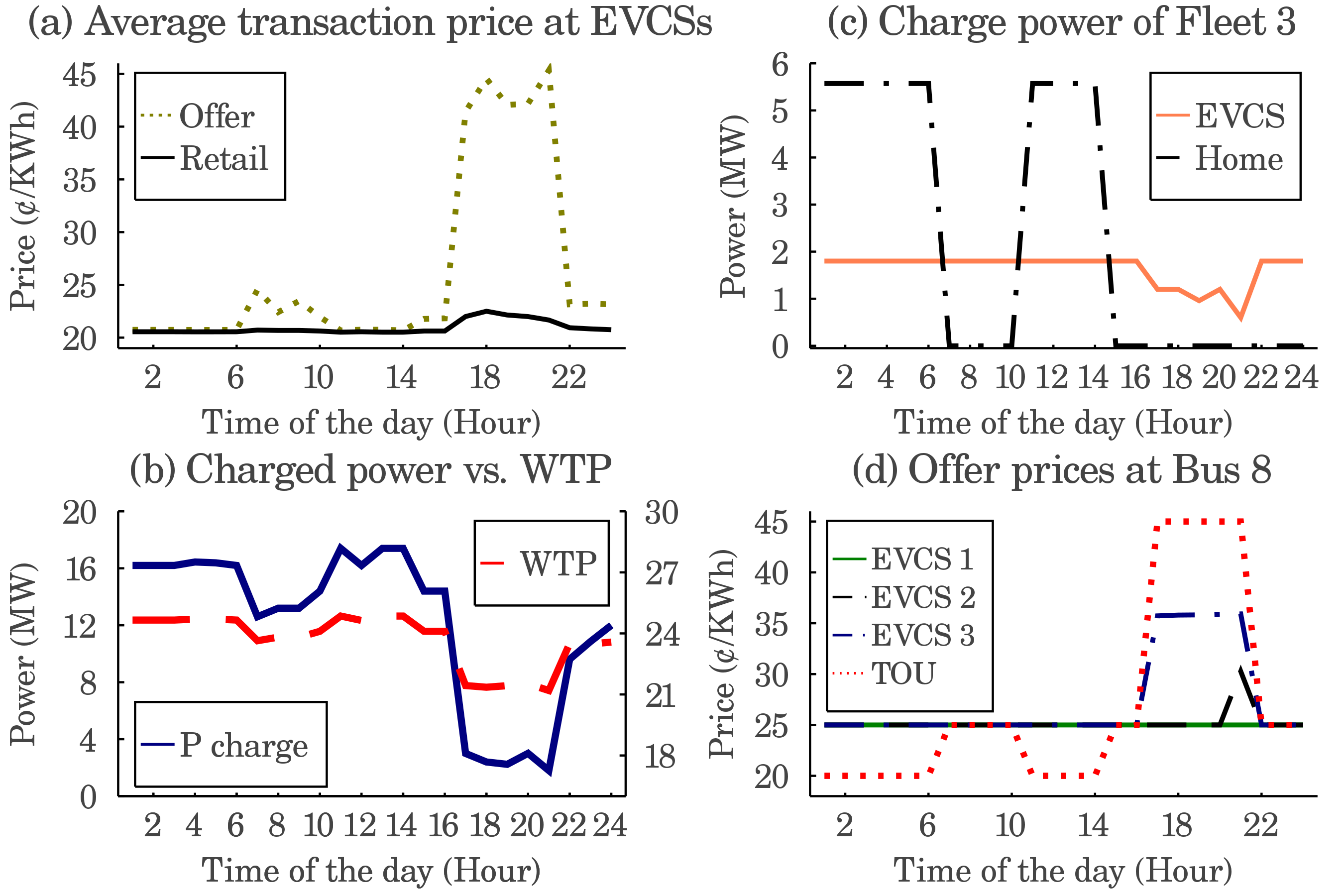}
    \vspace{-.75cm}
    \caption{Bidding, offering, and charged energy trends at the network}
    \label{fig:evcs_result_sell_charge}
    \vspace{-.35cm}
\end{figure}

In this case, the economic viability of the proposed structure with an EV penetration level of 10\% is investigated. It is observed that the participation of EVCSs in DAM is highly profitable to them. According to the results for the daily operation of the test system displayed in Fig. \ref{fig:evcs_result_sell_charge}, EVCSs have collectively accumulated \$55,621 in profits. EVCSs have sold an aggregated amount of \$60,596 to the customers, while they only paid \$4,974 for the purchased power in DAM, which is equal to a profit margin ratio of 11.35.

\par EV owners also gain from the market participation of EVCSs. In the presence of EVCSs, the combined EV owner charge of cost at home and EVCS locations is equal to \$81,697. While in the case where no EVCSs are present, EV owners are charged a total of \$88,030. This shows that besides improved public equity and accessibility which are immediate benefits of developing EVCSs in the system, total payments of EV owners are also reduced.

\par An interesting observation in Fig. \ref{fig:evcs_result_sell_charge} is the difference between the average offer price and the cleared retail price (total customer payments divided by total customer charged power). In Fig. \ref{fig:evcs_result_sell_charge} (a), the average offer price of EVCSs in the network peaks at 45.3 (\textcent/KWh). However, EVs tend to charge lower during periods with higher offer prices, causing the retail price to peak only at 22.5 (\textcent/KWh). WTP of EVCSs is correlated with the charged power at EVCS location. Fig. \ref{fig:evcs_result_sell_charge} (b) illustrates that in periods with higher charged power, WTP is higher and WTP drops when charged power reaches its minimum. It is observed in Table \ref{table:evcs_result_base} that at locations with higher retail prices, EVCSs tend to offer lower prices and bid higher. This strategy promotes revenues of EVCSs and lowers offer prices, which leads to increased social welfare and reduces customer payments. EVs schedule their charging to take place at home only when offer prices at EVCSs are higher than TOU rate and ECVSs have reached their maximum capacity. In parts (c) and (d) of Fig. \ref{fig:evcs_result_sell_charge} it is observed that during hours 7-10, even though the offer price at EVCSs is the same as TOU rate, EVs prefer to become charged at EVCS.

\begin{table}[h!]\vspace{-.35cm} \centering \caption{Offer prices and WTP bids and charged power comparison} \label{table:evcs_result_base}
\begin{tabular}{ccccc} \hline  \hline 
 & Bus 2 & Bus 4 & Bus 19 & Bus 29\\ \hline 
Average EVCS offer (\textcent/KWh) & 21.6 & 28.6 & 26.6 & 32.7 \\
Average EVCS WTP (\$/MWh) & 24.4 & 23.2 & 23.8 & 21.7 \\
Total charged at EVCSs (MWh) & 37.2 & 24.1 & 30.1 & 7.6 \\
Total charged at home (MWh) & 31.6 & 0 & 0 & 0 \\\hline \hline
\end{tabular}\vspace{-.55cm}
\end{table}

\par
\begin{figure}[h!]
    \centering
    \vspace{-.35cm}
   \includegraphics[width=\linewidth]{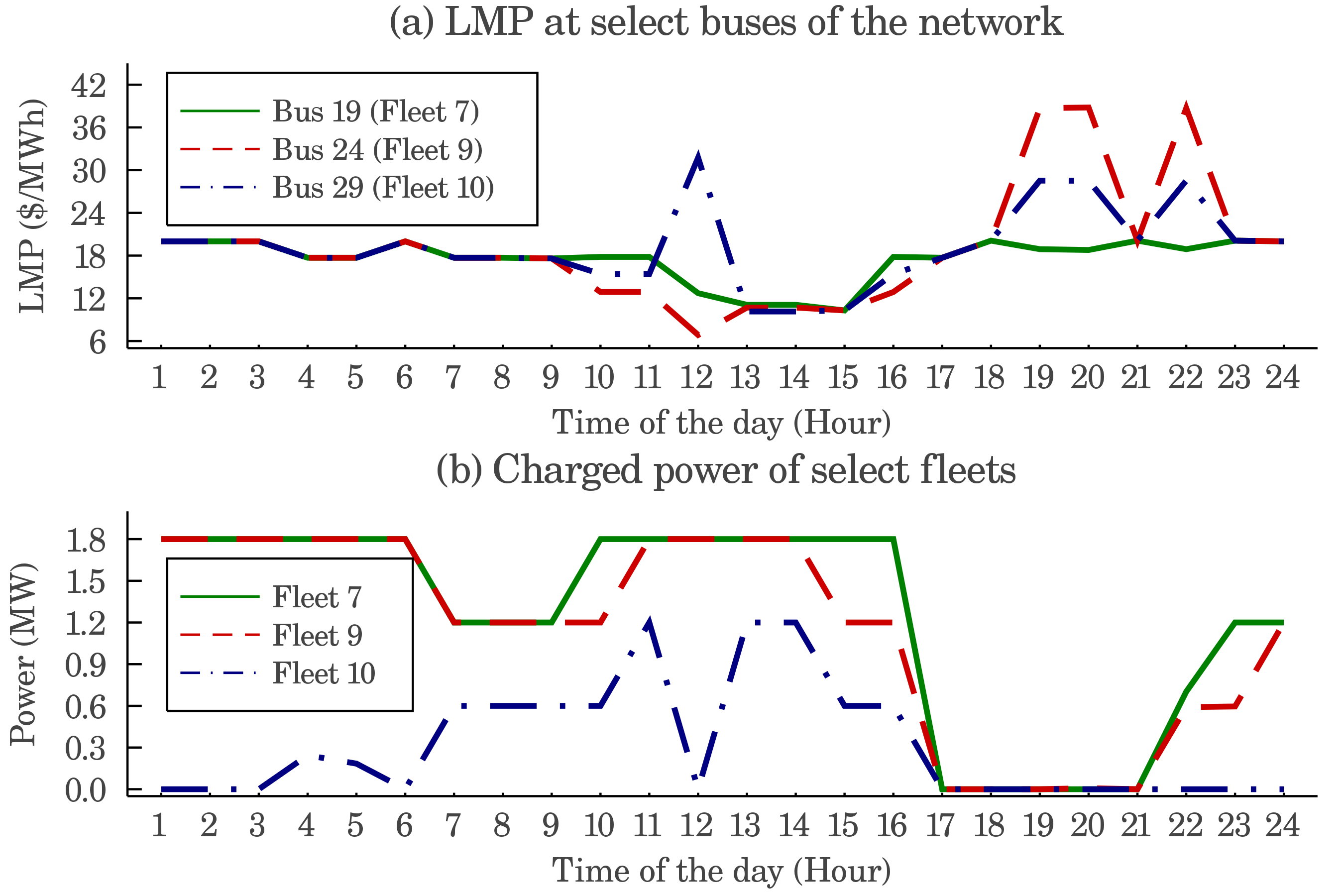}
    \vspace{-.75cm}
    \caption{LMP and charged power curve at different buses of the network}
    \label{fig:evcs_result_lmp} \vspace{-.45cm}
\end{figure}

 Another strategy of EVCSs is to schedule charging of fleets during periods when LMP is lower. The curves for LMP and charged power of fleets at select buses of the network are presented in Fig. \ref{fig:evcs_result_lmp}. 
It is observed that during the period 17-21 when LMPs are soaring on buses 24 and 29, the scheduled charge powers for the fleets connected on these buses (fleets 9 and 10, respectively) are 0. Conversely, the highest charging activity of fleets coincides with the period 11-14 in the noon when LMPs are lowest.

\subsection{Analyzing the Impact of EV Penetration Level}
In the next case, we explored the implications of higher EV penetration levels. According to the current vehicle electrification rates, it is only natural to explore the consequences of more EV penetration rates on the proposed structure. EV penetration level is defined as the total amount of EV charge demand divided by the overall system demand, including EV demand. The results of the simulations for various EV penetration levels are displayed in Table \ref{table:evcs_compare_penetration}.

\begin{table}[h!]\vspace{-.25cm} \centering \caption{Comparison of different EV penetration levels} \label{table:evcs_compare_penetration}
\begin{tabular}{ccccc} \hline  \hline 
EV Penetration level (\%) & 10 & 15 & 20 & 25\\ \hline 
Revenue of EVCSs ($\times\$1000$) & 60.6 & 93.1 & 132.6 & 171.4 \\
Cost of EVCSs ($\times\$1000$) & 4.9 & 8.3 & 14.4 & 94.68 \\
Profit of EVCSs ($\times\$1000$) & 55.6 & 84.8 & 118.2 & 76.7 \\
Profit percentage (profit/cost$\times100\%$) & 1135 & 1022 & 821 & 81 \\
Retail price (\textcent/KWh) & 20.6 & 20.6 & 20.7 & 20.3 \\
EVCSs purchased price (\$/MWh) & 16.7 & 18.4 & 22.5 & 122.2 \\
Network maximum LMP (\$/MWh) & 39.3 & 138.8 & 140.23 & 13219.0 \\
Network minimum LMP (\$/MWh) & 0 & 0 & 0 & -183.9 \\\hline \hline
\end{tabular}\vspace{-.25cm}
\end{table}

\par It is observed that increased EV penetration level does not necessarily translate into more profits for EVCSs. According to Table \ref{table:evcs_compare_penetration}, with the increase in EV penetration, the retail price almost remains the same, and EVCSs' revenues increase proportionally. However, it is observed that the cost of EVCSs is escalating very steeply. It is noticed that the LMP of the network is highly sensitive to the EV penetration, with the maximum LMP bursting from \$39 to \$13,219 with 15\% increased EV penetration. With the increased demand, it is only expected that more parts of the system be subject to congestion. EVCSs are charged by LMP price and that is why their purchased energy price grows substantially with more EV load.

\subsection{Analyzing the Impact of Solar Generation Installation}
 Studying renewable generations is important from two aspects. First, the number of renewable installations is rapidly increasing. Second, renewable outputs are unpredictable and dependent on weather conditions. In this case, the impact of solar generation installations was investigated by considering three scenarios with 0, 160 MW, and 320 MW installed PV unit capacity. 

\par According to the results displayed in Table \ref{table:evcs_compare_pv}, more PV installments lead to more profits for the EVCSs. With higher renewable penetration, the cost of electricity  decreases in the network. When no PV units are installed, the average energy purchase price is 19.8 (\$/MWh), whereas when 320 MW PV capacity is installed, the cost of energy is reduced by 42\%. However, it is noticed that since the retail price of EVCSs is not affected by the presence of PV units, they can make more profit. It is interesting to notice that in the case with 320 MW installed solar capacity, around 32\% of renewable outputs are curtailed. However, the operation of EVCSs is not affected by this issue.

\begin{table}[h!] \vspace{-.45cm} \centering \caption{Comparison of different solar generation installed capacity} \label{table:evcs_compare_pv}
\begin{tabular}{ccccc} \hline  \hline 
Installed PV capacity (MW) & 0 (0\%) & 160 & 320 \\ \hline 
Revenue of EVCSs ($\times\$1000$) & 59.5 & 60.6 & 60.0 \\
Cost of EVCSs ($\times\$1000$) & 5.71 & 4.9 & 3.36 \\
Profit of EVCSs ($\times\$1000$) & 53.7 & 55.6 & 56.7 \\
Profit percentage (profit/cost$\times100\%$) & 940 & 1135 & 1688 \\
Retail price (\textcent/KWh) & 20.6 & 20.7 & 20.6 \\
EVCSs purchased price (\$/MWh) & 19.8 & 16.7 & 11.5 \\
Total solar utilized (MWh) & 0 & 1113.8 & 1518.1 \\
Solar curtailments (\%) & 0 & 0.4 & 32.1 \\
Network maximum LMP (\$/MWh) & 38.7 & 39.3 & 38.8 \\
Network minimum LMP (\$/MWh) & 12.3 & 0 & 0 \\\hline \hline
\end{tabular} \vspace{-.45cm}
\end{table}

\section{Conclusion and Future Works}
In this work, a bidding/offering strategy in competition with regulated time-of-use pricing is proposed for EVCSs participating in DAM to investigate their profitability. According to the case study results, EVCSs can successfully earn profits by purchasing energy in DAM and selling it to EV owners. It was shown that the customers also benefit from the proposed structure, and the overall EV owner payments is reduced compared with the case where no EVCSs are present in the system. Additional cases are dedicated to analyzing the impact of different EV and renewable penetration rates. It was shown that EVCSs make less profit with more EV penetration and more profit with more PV unit installments. Possible continuations to this study include investigating the proposed structure in the long-term planning period and analyzing competitive strategies between different EVCSs.

\bibliographystyle{IEEEtran}

\bibliography{ev_charge_sched.bbl}

\end{document}